\documentstyle[prl,aps,array,epsfig,multicol]{revtex}

\draft
\begin{document}
\title{Wave scattering from self-affine surfaces}
\author{Ingve Simonsen$^{*\dag}$, Damien Vandembroucq$^{*}$
 and St\'ephane Roux$^{*}$}
\address{
$^*$Laboratoire CNRS/Saint-Gobain ``Surface du Verre et Interfaces'', 
    93303 Aubervilliers Cedex, France\\
$^\dag$Department of Physics, 
       The Norwegian University of Science and Technology,
       N-7491 Trondheim, Norway}
\date{\today}
\maketitle

\begin{abstract} 
  Electromagnetic wave scattering from a perfectly reflecting
  self-affine surface is considered.  Within the framework of the
  Kirchhoff approximation, we show that the scattering cross section
  can be exactly written as a function of the scattering angle via a
  centered symmetric L\'evy distribution for general roughness
  amplitude, Hurst exponent and wavelength of the incident wave. The
  amplitude of the specular peak, its width and its position are
  discussed as well as the power law decrease (with scattering angle)
  of the scattering cross section.
\end{abstract}

\begin{multicols}{2}


Wave scattering from rough surfaces has been studied for long
\cite{Rayleigh,Beckmann} (see also \cite{Ogilvy,Nieto,Voro} for recent
reviews of the subject) with potential applications in remote sensing,
acoustical/optical/radar detection or design of surfaces with
specified scattering properties... The theoretical prediction of the
angular distribution of the scattered intensity requires a proper
statistical description of the surface roughness. At this stage it is
classical to assume that the statistics of the height and its
correlation function are Gaussian.  This assumption allows to build
analytical expressions and to compare them with experimental or
numerical results. 

Since the pioneering work of Mandelbrot\cite{Mandelbrot}, however,
scale invariance has emerged as a relevant tool to describe the
geometry of real ``disordered'' objects. In the case of surfaces, the
scale invariance takes the form of self-affinity. Such surfaces remain
invariant under dilation of different ratios over the horizontal and
the vertical directions. This long range correlated height
distribution is characterized by a roughness exponent, an amplitude
parameter, and by the lower and upper limits of the scaling invariance
region (instead of a height standard deviation and a two point
correlation function of finite width).  It turned out that many real
surfaces can be described through this formalism. Surfaces obtained
from fracture \cite{MP,EB,Maloy}, growth or deposition processes
\cite{Meakin} are classical examples.  Similar results have
been more recently found for surfaces obtained from cold metal
rolling\cite{laminage}.  Note that such scaling invariance properties
are exhibited from nanometric scales for native float glass
\cite{Creuzet} up to kilometric scales for geological faults
\cite{Schmittbuhl}.

The aim of this letter is to exhibit an analytical expression of the
distribution of the scattering cross section for a perfectly
reflecting self-affine surface in the framework of the Kirchhoff
approximation.  Since a first paper examining the effect of scale
invariance in the context of scattering from rough
surfaces~\cite{Berry}, a large amount of studies have been published
in various journals (see for example
Ref.\cite{Jaggard,Shepard,MacSharry,Lin,Chen,Sheppard,Sanchez,Sanchez2,Zhao}).
Two main points motivated most of these studies {\it i)} the effect of
scale invariance on the field scattered from the surface in comparison
with the more classical case of a finite width correlation length {\it
  ii)} the hope of directly measuring the ``fractal'' parameters of
the surface {\it via} an acoustical or optical experiment.  In spite
of a large amount of works, it is fair to say that concerning far
field scattering the only stable result that emerged is the existence
of a power law tail in the intensity distribution with an exponent
directly related to the roughness exponent. Very few analytical
predictions can be found and oppositely a large number of numerical
simulations still lack a clear physical interpretation.

A reason for this could be an
excessive focus on the ``fractal dimension'' (conversely the roughness
exponent) with a blind point on the other parameters (the limits of
the scale invariance regime and more often the amplitude parameter).  A
noteworthy exception is, however, due to Jakeman and his
collaborators\cite{Jakeman86,Jakeman88} who predicted a L\'evy
distribution for the intensity of a wave scattered from a self-affine
random phase screen. However, these authors did not extend these
analytical results to the case of scattering from surfaces.

In the following we briefly recall the main properties of self-affine
surfaces and we use the Kirchhoff approximation to show that the 
scattering cross section can be written as a function of the
scattering angle via a centered symmetric L\'evy distribution.

A surface is  self-affine between the scales $\xi_{-}$
and $\xi_{+}$ if it remains (either exactly or
statistically) invariant in this region under
transformations of the form:
\begin{equation}
\left(x,y,z \right) \to 
\left(\lambda x,\lambda y,\lambda^{H} z \right).
\end{equation}
The exponent, $H$, is usually called the roughness or Hurst exponent.
Restricting ourselves to profiles, a statistical translation of the
previous statement is that the probability $p(\Delta z;\Delta x)$ of
having a height difference $\Delta z$ over the distance $\Delta x$ or
its cumulative ${\cal P}(\Delta z;\Delta x)=\int_{-\infty}^{\Delta z}
p(\delta z;\Delta x) d \delta z$ is such that: 
\begin{eqnarray}
{\cal  P}\left(\Delta z ; \Delta x \right) 
     = {\cal P}\left(\lambda^{H} \Delta z; \lambda \Delta x \right) 
=\Phi\left( \frac{\Delta z}{\Delta x^H}\right). 
\end{eqnarray}
Simple algebra based on this scaling gives
\begin{eqnarray}
\sigma \left( \Delta x \right) = \ell^{1-H} \Delta x^{H}\;,
\end{eqnarray}
where $\sigma$ is the standard deviation of the height profile.  Here
$\ell$ denotes a length scale, also known as the {\it topothesy}.
This quantity is defined by \protect$\sigma(\ell)=\ell$, which
allows the geometrical interpretation of the topothesy as the length
scale over which the profile has a mean slope of 45 degrees.  The
smaller $\ell$, the flatter the profile appears on a macroscopic
scale.  In the case of a Gaussian height distribution, the probability
$p(\Delta z;\Delta x)$ reads:
\begin{eqnarray}
\label{dist}
p(\Delta z;\Delta x)= \frac{1}{\sqrt{2\pi} \ell^{1-H} \Delta x^{H}}
\exp \left[ -\frac{1}{2} \left( 
\frac{\Delta z} {\ell^{1-H} \Delta x^{H}}
\right)^2 \right].
\end{eqnarray}
The self-affine profile is thus fully characterized by its exponent
$H$, its topothesy parameter $\ell$ and the bounds of the self-affine
regime $\xi_{-}$ and $\xi_{+}$.  Let us mention in addition that the
above geometric interpretation of the roughness amplitude made above
{\em does not} imply that $\ell$ has to lie between the
lower~($\xi_{-}$) and upper~($\xi_{+}$) cut-off for the self-affine
regime.  For the surfaces usually considered in scattering problems,
we rather expect that $\ell \ll \xi_{-}$. When $\xi_{-} < \ell <
\xi_{+}$, the topothesy makes the transition between the scales, below
$\ell$, for which a fractal dimension $D=2-H$ can be measured using
the box counting method and the scales, above $\ell$, for which this
dimension is just unity.

\begin{figure}
\begin{center}
\epsfig{file=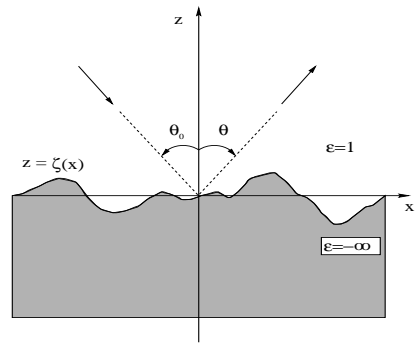,angle=0,width=0.3\textwidth,height=0.25\textwidth}
\end{center}
\caption{Sketch of the scattering geometry.}
\label{Fig:1}
\end{figure}

In the following we consider the scattering of s-polarized
electromagnetic waves from a one dimensional, random, self-affine
surface. This surface, of Hurst exponent $H$, we denote by
$z=\zeta(x)$, and it is assumed to be Gaussian self-affine.  We will
further assume that the lower limit of the self-affine regime is
smaller than the wavelength, $\lambda$, of the incident wave.  The
scattering geometry considered is depicted in Fig.~\ref{Fig:1}. The
incident plane is assumed to be the $xz$-plane, and the rough surface,
which is perfectly conducting, is illuminated from the vacuum side by
a plane wave of frequency $\omega=2\pi/\lambda$.  The incident and
scattering angle respectively we denote by $\theta_0$ and $\theta$,
and they are defined positive according to the convention indicated in
Fig.~\ref{Fig:1}.  In the above scattering geometry there is no
depolarization, and the electromagnetic field is represented by the
single non-vanishing component of the electric field
$\Phi(x,z|\omega)=E_y(x,z|\omega)$, which should satisfy the (scalar)
Helmholtz equation:
\begin{eqnarray}
    \label{Wave-eq}
    \left(\partial^2_x + \partial^2_z + \frac{\omega^2}{c^2}
    \right)\Phi(x,z|\omega) &=& 0,
\end{eqnarray}
with vanishing boundary condition on $z=\zeta(x)$ and outgoing 
wave condition at infinity.
In the far field region, above the surface, the field can be
represented as the sum of an incident wave and scattered waves:
\begin{eqnarray}
    \label{asymp-form}
    \Phi(x,z|\omega) &=& \Phi_0(x,z|\omega)
      +  \int^{\infty}_{-\infty}\; \frac{dq}{2\pi} 
           R(q|k)\; e^{iqx+i\alpha_0(q,\omega)z},
\end{eqnarray}
where the plane incident wave is given by:
\begin{eqnarray}
    \label{inc-field}
    \Phi_0(x,z|\omega) &=& \exp\left\{ikx-i\alpha_0(k,\omega)z\right\}
\end{eqnarray}
and $R(q|k)$ is the {\em scattering amplitude}. In the above
expression, we have defined
$\alpha_0(q,\omega)=\sqrt{(\omega/c)^2-q^2}$
(${\Re}\alpha_0(q,\omega)>0$, ${\Im}\alpha_0(q,\omega)>0$).
Furthermore, the momentum variables $q$ and $k$ are in the
radiative region related to respectively the scattering and incident angle
by $q=(\omega/c)\sin\theta$ and $k=(\omega/c)\sin\theta_0$, 
so that $\alpha_0(q,\omega)=(\omega/c)\cos\theta$ and
$\alpha_0(k,\omega)=(\omega/c)\cos\theta_0$.

Our aim is to obtain an expression for the scattering amplitude, since
this quantity determines the scattered field. The mean differential
reflection coefficient (or mean scattering cross section), which is an
experimentally accessible quantity, and defined as the fraction of the
total, time-averaged, incident energy flux scattered into the angular
interval $(\theta,\theta+d\theta)$, is related to this quantity by the
following expression~\cite{AnnPhys}:
\begin{eqnarray}
    \label{MDRC}
    \left< \frac{\partial R}{\partial \theta} \right>
    &=& \frac{1}{L}\;\frac{\omega}{2\pi c}\;\frac{\cos^2\theta}{\cos\theta_0}
      \left< \left| R(q|k)\right|^2\right>.    
\end{eqnarray}
Here $L$ denotes the length of the self-affine profile function as
measured along the $x$-direction, and the other quantities have been
defined earlier.  The angle brackets denote an ensemble average over
the surface profiles $\zeta(x)$, and the momentum variables are
understood to be related to the angles $\theta_0$ and $\theta$
according to the expressions given above. 

\narrowtext

\begin{figure}
\begin{center}
  \epsfig{file=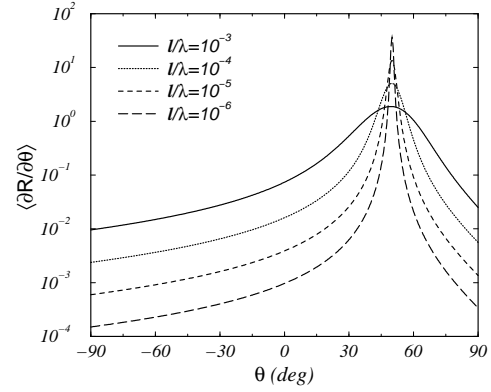,angle=0,width=0.35\textwidth}
\end{center}
\caption{The mean DRC as defined by Eq.~(\ref{final-result}) for an
incidence angle of $\theta_0=50^\circ$ and a roughness exponent of
$H=0.7$. The surfaces had topothesy over wavelength of
$\ell/\lambda=10^{-3}$, $10^{-4}$, $10^{-5}$ and $10^{-6}$ which
gives typical slopes at the wavelength scale of respectively 
$s(\lambda)=0.126$, $0.063$, $0.032$ and $0.016$.}
\label{intensity}
\end{figure}

We now use the Kirchhoff-approximation, which consists of locally
replacing the surface by its tangential plane, and thereafter using
the (local) Fresnel reflection coefficient for the local incident
angle to obtain the scattered field.  Notice here that dealing with a
surface whose scaling invariance range is bounded by a lower cut-off
$\xi_{-}$ does ensure that the tangential plane is well defined in
every point.  The scattering amplitude is then~\cite{AnnPhys}:
\begin{eqnarray}
    \label{scattering-amp}
    R(q|k) &=& \frac{-i}{2 \alpha_0(q,\omega)} 
     \int^{L/2}_{-L/2}dx \;
     e^{-iqx-i\alpha_0(q,\omega)\zeta(x)} {\cal N}(x|\omega),
\end{eqnarray}
where $ {\cal N}(x|\omega)$ is a source function defined by 
${\cal N}(x|\omega)=2\,\partial_n
\!\!\left.\Phi_0(x,z|\omega)\right|_{z=\zeta(x)}$, with the
(unnormalized) normal derivative  given by
$\partial_n=-\zeta'(x)\partial_x+\partial_z$.  By substituting this
expression for the source function into Eq.~(\ref{MDRC}), one can show
after some straightforward algebra where one takes advantage of the
self-affine profile function having stationary increments, that the
mean differential reflection coefficient is given by
\begin{eqnarray}
    \label{MDRC-mod}
    \left< \frac{\partial R}{\partial \theta} \right>
    &=&
    \frac{\omega}{2\pi c} \frac{1}{\cos\theta_0}
    \frac{\cos^2\frac{\theta+\theta_0}{2} }{
    \cos^2\frac{\theta-\theta_0}{2}} \nonumber \\
    & &  \times 
     \int^{L/2}_{-L/2}dv\; 
         \exp\left\{
                 i\frac{\omega}{c}(\sin\theta-\sin\theta_0)v
             \right\} \Omega(v) ,
\end{eqnarray}
where 
\begin{eqnarray}
    \label{Omega} 
    \Omega(v) &=& \left< \exp\left\{-i\frac{\omega}{c}[\cos\theta+\cos\theta_0]\,
                                     \Delta\zeta(v)\right\}\right>,
\end{eqnarray}
with $\Delta\zeta(v)=\zeta(x)-\zeta(x+v)$.
Note that the statistical properties of the profile function,
$\zeta(x)$, enters Eq.~(\ref{MDRC-mod}) only through $\Omega(v)$.
With Eq.~(\ref{dist}), one may now calculate the ensemble average
contained in $\Omega(v)$. For a Gaussian self-affine surface one gets:
\begin{eqnarray}
    \label{Omega-2} 
    \Omega(v) &=& 
      \int^{\infty}_{-\infty} dz\;
      \exp\left\{ -i\frac{\omega}{c}\left(
      \cos\theta+\cos\theta_0\right)z\right\}
          p(z; v)
      \nonumber \\
      &=&  
      \exp\left\{ -\left(\frac{\omega}{c} 
                     \frac{\cos\theta+\cos\theta_0}{\sqrt{2}}
                       \ell^{1-H}v^H
                   \right)^2
             \right\}.
\end{eqnarray}
By performing the change of variable:
\begin{eqnarray}
    \label{s-var} 
u &=& v \left[ \frac{\omega}{c} 
                       \frac{\cos\theta+\cos\theta_0}{\sqrt{2}} \ell^{1-H}
        \right]^{1/H} \;,
\end{eqnarray}
in Eq.~(\ref{MDRC-mod}), and letting the length of the profile extent
to infinity, $L\rightarrow \infty$, one finally obtains the following
expression for the mean differential reflection coefficient:
\begin{mathletters}
    \label{final-result}
\begin{eqnarray}
    \label{MDRC-final}
    \left< \frac{\partial R}{\partial \theta} \right>
    &=&
    \frac{a^{-(\frac{1}{H}-1)}}{\sqrt{2}\, \cos\theta_0 }
     \frac{
    \cos\frac{\theta+\theta_0}{2}}{\cos^3\frac{\theta-\theta_0}{2}}  
 {\cal L}_{2H}\left(\frac{ \sqrt{2}\tan\frac{\theta-\theta_0}{2} }{ a^{\frac{1}{H}-1}} \right),  
\end{eqnarray}
where 
   $a = 
   \sqrt{2}\frac{\omega}{c}\ell
        \cos\frac{\theta+\theta_0}{2}\cos\frac{\theta-\theta_0}{2}$,
and ${\cal L}_{\alpha}(x)$ is the centered symmetrical L\'evy stable
distribution of exponent $\alpha$ defined as
\begin{eqnarray}
    \label{Levy-distribution-1}
    {\cal L}_\alpha(x) &=& \frac{1}{2\pi} \int^\infty_{-\infty} dk \; 
                 e^{ikx} e^{-\left| k \right|^\alpha}. 
\end{eqnarray}
\end{mathletters}
Note that in the above expressions the wavelength, $\lambda$, only
comes into play through the combination
$(\ell/\lambda)^{1-\frac{1}{H}}$ which appears both in the prefactor
and in the argument of the L\'evy distribution. This quantity can be
geometrically regarded as $s(\lambda)^{1/H}$ where
$s(\lambda)=(\ell/\lambda)^{1-H}$ is the typical slope of the surface
over a wavelength. The behavior of the scattered intensity is thus
entirely determined by this typical slope $s(\lambda)$ and the
roughness exponent H.  Figure \ref{intensity} shows the mean
differential reflection coefficient~(DRC), obtained for a incident
angle of $\theta_0=50^\circ$ for different values of the ratio
$\ell/\lambda$.  
Note that the L\'evy distribution obtained above is likely not to be specific 
of the Kirchhoff approximation: the integrand in (\ref{MDRC-mod}) is indeed
nothing but the reflection of the statistical translational invariance
of the problem.
Taking advantage of the asymptotic expansion of the L\'evy
distribution around zero\cite{Levyzero} we find that the
amplitude of the specular peak scales as
\begin{eqnarray}
\left. \left< \frac{\partial R}{\partial \theta} \right>
\right|_{\theta=\theta_0}
\simeq 
\frac{\Gamma(\frac{1}{2H}) }
   {2\sqrt{2}\pi H \left(2\sqrt{2}\pi\frac{\ell}{\lambda} \cos\theta_0 
   \right)^{\frac{1}{H}-1} }, 
\end{eqnarray}
and its  half width
 (assuming
$\frac{\omega}{c}\ell\cos\theta_0\ll 1$) as
\begin{eqnarray}
w &\simeq&  2 \sqrt{ 
     \frac{\Gamma(\frac{1}{2H})}{\Gamma(\frac{3}{2H})} }
   \left( 
        2\sqrt{2}\pi\frac{\ell}{\lambda} \cos\theta_0 
   \right)^{\frac{1}{H}-1 }.
\end{eqnarray}
Notice that in the case of a non zero angle of incidence, $\theta_0$,
this (specular) peak is located at $\theta=\theta_0 + \Delta\theta_0$,
where
$\Delta\theta_0$~($\Delta\theta_0 \sim w^2\ll w$) scales as
\begin{eqnarray}
\Delta\theta_0 \simeq 
 \frac{2H-1}{H} \frac{\Gamma(\frac{1}{2H})}{\Gamma(\frac{3}{2H})}
   \tan\theta_0 \left( 2\sqrt{2}\pi\frac{\ell}{\lambda}\cos\theta_0 
                 \right)^{\frac{2}{H}-2},
\end{eqnarray}
i.e. we have, due to the self-affinity of the surface, a shift,
$\Delta\theta_0$, of the specular direction as compared to its
``classical'' position $\theta=\theta_0$.

It is interesting that these non-trivial scaling results can all be
retrieved via simple dimensional arguments. Let us examine the
intensity scattered in direction $\theta$; in a naive Huyghens
framework two different effects will compete to destroy the coherence
of two source points on the surface {\it i)} the angular distance
separating $\theta$ from the specular direction {\it ii)} the
roughness. The two source points 
will interfere coherently if  the distance between them is lower than the 
 two length scales 
$d_{ang}=\lambda / p$ where $p=\tan[(\theta-\theta_0)/2]$ is the slope
of the isophase plane and  $d_{rough}$,  the
typical length scale over which the height difference is of order
$\lambda $.
 In case of a self-affine surface,
$d_{rough}=\lambda^{\frac{1}{H}} \ell^{1-\frac{1}{H}}$. Depending on
the observation angle $\theta$, the coherence will be controlled
either by $d_{rough}$ or by $d_{ang}$, which will correspond
respectively to the peak or the tail of the intensity distribution.
The width $w$ of the peak can then be defined as the transition
between these two regions and will be such that $d_{rough}\simeq
d_{ang}$.  This leads directly to $w \propto ( {\ell}/{\lambda}
)^{\frac{1}{H}-1}$.  While discussing the width of the peak present in
the mean DRC, we also have to take into account the fact that real
self-affine surfaces are not scale invariant over an infinite range of
length scales. The upper cut-off $\xi_{+}$ introduces a convolution of
the preceding results by a function of width $\lambda/\xi_{+}$. This
convolution does not affect the tails of the distribution but
considering the peak we have now to distinguish two regimes whose
transition will be defined by $\sigma=\ell^{1-H}\xi_{+}^H \simeq
\lambda$ where $\sigma$ denotes here the standard deviation of the
height profile measured over a macroscopic scale {\it i.e.} larger
than $\xi_{+}$. For $\sigma < \lambda$ the width of the peak will be
$w \simeq (\ell/\lambda )^{\frac{1}{H}-1} \simeq
s(\lambda)^{\frac{1}{H}}$ and for $\sigma > \lambda$ we will get $w
\simeq \lambda/\xi_{+}$.

Using now the expansion of the L\'evy distribution at
infinity\cite{Levyinf} we get for the diffuse intensity
($\theta\neq \theta_0$):
\begin{eqnarray}
\nonumber
\left< \frac{\partial R}{\partial \theta}\right> &\simeq&
   \frac{\Gamma(1+2H) \sin(\pi H)}{4 \pi}   
        \frac{    \left(4\pi\frac{\ell}{\lambda}\right)^{2-2H} 
           \left(\cos\frac{\theta+\theta_0}{2}\right)^{3-2H}}{ 
        \cos\theta_0 \left| \sin\frac{\theta-\theta_0}{2}\right|^{1+2H}} \;.
\end{eqnarray}
The dependence of the prefactor on the surface
parameters $\ell$ and $H$ is a new result of this letter. 
If the scaling can be easily retrieved {\it via} a first order
expansion in plane waves, the above expression suggests that provided
single scattering is considered, the power law behavior survives even
in a non perturbative approach. 


We proposed a direct derivation of the intensity distribution of a
wave scattered from a perfectly conducting self-affine rough surface
in the framework of single scattering. Using a simple Kirchhoff
approximation we obtain an exact solution consistent with all
previously published perturbative or numerical results.
Among other examples, our results could be directly tested in the case
of an acoustic wave scattered from a fracture surface. 
This work will be extended to deal with more complicated problems
directly related to scattering from rough surfaces (shadowing effects,
multiple scattering, dielectric media...).  Numerical investigations
about the range of validity of the results obtained in this letter
will follow.

\medskip

The authors would like to thank Claude Boccara, Jean-Jacques Greffet,
Tamara Leskova and Alexei A. Maradudin for useful comments about this
work.  I.S. would like to thank the Research Council of Norway
(Contract No. 32690/213), Norsk Hydro ASA, and Total Norge ASA for
financial support.


\end{multicols}
\end{document}